\documentclass[a4paper,11pt]{article}
\usepackage{pos}
\usepackage[T1]{fontenc}
\usepackage[utf8]{inputenc}
\usepackage[english]{babel}
\usepackage{amsmath}
\usepackage{graphicx}
\usepackage{color}
\usepackage{transparent}
\usepackage[percent]{overpic}
\usepackage{mciteplus}
\usepackage{todonotes}
\usepackage[normalem]{ulem}
\usepackage{lineno}
\renewcommand{\i}{\text{i}}
\renewcommand{\d}{\text{d}}
\renewcommand{\O}{\mathcal{O}}
\newcommand{\LQCD}{\Lambda_{\text{QCD}}}
\newcommand{\als}{\alpha_{\text{s}}}
\newcommand{\muus}{\mu_{\text{us}}}
\newcommand{\CF}{C_{\text{F}}}
\newcommand{\Nf}{N_{\text{f}}}
\newcommand{\mc}{m_{\text{c}}}
\newcommand{\MS}{\overline{\text{MS}}}
\newcommand{\tr}{\ensuremath{\mathop{tr}}}
\let\oldthebibliography\thebibliography
\renewcommand\thebibliography[1]{%
\oldthebibliography{#1}%
\setlength{\parskip}{.2\baselineskip plus .05\baselineskip minus .05\baselineskip}%
\setlength{\itemsep}{0pt plus 0.3ex}%
}
\let\todoorig\todo\renewcommand\todo[1]{\todoorig[inline]{#1}} 
\title{The static energy in \texorpdfstring{$2+1+1$}{2+1+1}-flavor QCD \hfill \normalsize TUM-EFT 153/21}
\author*[a]{Sebastian Steinbeißer}
\author[a,b,c]{Nora Brambilla}
\author[d]{Rafael L.~Delgado}
\author[e]{Andreas~S.~Kronfeld}
\author[a]{Viljami Leino}
\author[f]{Peter Petreczky}
\author[a]{Antonio Vairo,}
\author[a,g]{Johannes~Heinrich Weber}
\affiliation[a]{Physik Department, Technische Universität München, James-Franck-Str.~1, D-85748 Garching b. München, Germany}
\affiliation[b]{Institute for Advanced Study, Technische Universität München, Lichtenbergstrasse 2a, D-85748 Garching b. München, Germany}
\affiliation[c]{Munich Data Science Institute, Technische Universität München, Walther-von-Dyck-Strasse 10, D-85748 Garching b. München, Germany}
\affiliation[d]{Universidad Politécnica de Madrid, Nikola Tesla, s/n, 28031-Madrid, Spain}
\affiliation[e]{Particle Theory Department, Fermi National Accelerator Laboratory, Batavia, IL 60510-5011, USA}
\affiliation[f]{Brookhaven National Laboratory, Upton, NY 11973-5000, USA}
\affiliation[g]{Institut für Physik \& IRIS Adlershof, Humboldt-Universität zu Berlin, Zum Großen Windkanal~6, D-12489 Berlin, Germany}
\emailAdd{sebastian.steinbeisser@tum.de}
\emailAdd{nora.brambilla@tum.de}
\emailAdd{rafael.delgado@upm.es}
\emailAdd{ask@fnal.gov}
\emailAdd{viljami.leino@tum.de}
\emailAdd{petreczk@bnl.gov}
\emailAdd{antonio.vairo@tum.de}
\emailAdd{dr.rer.nat.weber@gmail.com}
\abstract{%
\vspace*{-6mm}
\textbf{\textsf{TUMQCD Collaboration}}\\[1em]
We report on the status of the analysis of the static energy in $2+1+1$-flavor QCD.
The static energy is obtained by measuring Wilson line correlators in Coulomb gauge using the HISQ action, yielding the scales $r_{0}/a$, $r_{1}/a$, $r_{2}/a$, their ratios, and the string tension~$\sigma  r_{i}^{2}$.
We put emphasis on the possible effects due to the dynamical charm-quark by comparing the lattice results to continuum results of the static energy with and without a massive flavor at two-loop accuracy.
We employ gauge-field ensembles from the HotQCD and MILC Collaborations.
}

\FullConference{%
 The 38th International Symposium on Lattice Field Theory, LATTICE2021
  26th-30th July, 2021
  Zoom/Gather@Massachusetts Institute of Technology
}
\begin{document}

\maketitle

\section{Introduction}\vspace{-1mm}

The static energy, $E_{0}(r)$, of an infinitely heavy quark-antiquark pair is an important observable that has been studied since the mid-1970s in both lattice gauge theory~\cite{Wilson:1974sk, Brown:1979ya} and perturbation theory~\cite{Fischler:1977yf, Billoire:1979ih}.
The perturbative formulas for $\Nf$ massless flavors are known at N$^{3}$LO, with logarithms known at N$^{4}$LO~\cite{Brambilla:1999qa_short, Anzai:2009tm, *Smirnov:2009fh, *Brambilla:2006wp_short}.
A compact summary of the perturbative expression of $E_{0}(r)$ can be found in~\cite{Tormo:2013tha}.
The comparison between lattice data and perturbation theory allows one to extract $\als$.
This has been done in the pure gauge SU(3) theory starting from the seminal work of~\cite{Necco:2001xg} (a more recent reference is~\cite{Brambilla:2010pp_short}) and, subsequently, in QCD with dynamical fermions.
The TUMQCD Collaboration has extracted $\als$ from the static energy computed in $2+1$-flavor lattice QCD since 2012~\cite{Bazavov:2012ka_short, Bazavov:2014soa_short, Bazavov:2019qoo_short}.
We are currently extending this program to $2+1+1$-flavor QCD.
Charm quark mass effects could be important at distances of the order of the inverse of the charm mass, $\mc$, where comparisons with perturbation theory are still sensible and affecting the determination of $\als$.
With this work we compare for the first time a $2+1+1$-flavor lattice calculation of the static energy with perturbative QCD including finite charm mass corrections at N$^{2}$LO accuracy.
We employ gauge-field ensembles from the HotQCD and MILC Collaborations in this work.
More details will be presented in a forthcoming publication~\cite{actual_paper}.

\section{Static energy and charm-quark effects in perturbation theory}

\subsection{Static quark-antiquark energy in perturbation theory}

The static energy, $E_{0}(r)$, is the energy stored in a static quark-antiquark pair, separated by a distance $r$.
It is related to the real-time Wilson loop~\cite{Wilson:1974sk, Susskind:1976pi, Fischler:1977yf, Brown:1979ya}:
\begin{equation}
E_{0}(r) = \lim\limits_{t \to \infty} \frac{\i}{t} \ln \left\langle \tr \mathcal{P} \exp\left\{ \i g \oint\limits_{r \times t} \d z^{\mu} A_{\mu}(z) \right\} \right\rangle ,
\end{equation}
where the integral is over a rectangle of spatial length $r$ and temporal length $t$, $\langle\dots\rangle$ denotes the expectation value, $\mathcal{P}$ stands for the path ordering of the color matrices, $g$ is the QCD gauge coupling ($\als = g^{2}/(4\pi)$), and $A_{\mu}$ are the SU(3) gauge fields.
The spatial Wilson lines can be omitted in a suitable gauge, e.g., in Coulomb gauge.

At short distances, where $r \LQCD \ll 1$, we have that $\als(1/r) \ll 1$ and $E_{0}(r)$ may be computed as a perturbative series in $\als$ of the form:
\begin{equation}
\label{eq:statenergyI}
E_{0}(r) = \Lambda - \frac{\CF \als}{r} \left( 1 + \# \als + \# \als^{2} + \# \als^{3} + \# \als^{3} \ln\als + \dots \right) ,
\end{equation}
where $\Lambda$ is a constant of mass dimension one.
Through two loops, the perturbative expansion depends only on a soft scale $\nu$ of order $1/r$.
Starting at three loops, however, another ultrasoft energy scale $\muus$ of order $\als/r$ affects the static energy.
Soft and ultrasoft effects may be conveniently factorized in an effective field theory framework~\cite{Brambilla:1999qa_short, Brambilla:1999xf_short},
\begin{equation}
\label{eq:statenergyII}
E_{0}(r) = \Lambda + V_{\text{s}}(r,\nu,\muus) + \delta_{\text{us}}(r,\nu,\muus) ,
\end{equation}
where $V_{\text{s}}(r,\nu,\muus)$ contains all soft contributions and can be identified with the color-singlet static potential (as the term is used in the context of perturbative theory), and $\delta_{\text{us}}(r,\nu,\muus)$ encodes the ultrasoft contributions.

The static energy is an observable and hence must be finite, but the functions $V_{\text{s}}(r,\nu,\muus)$ and $\delta_{\text{us}}(r,\nu,\muus)$ not necessarily so.
Indeed, the $\ln\als$ terms appearing in the expansion~\eqref{eq:statenergyI} starting from N$^{3}$LO are remnants of the cancellations between infrared divergences in the soft function $V_{\text{s}}(r,\nu,\muus)$ and ultraviolet divergences in the ultrasoft function $\delta_{\text{us}}(r,\nu,\muus)$.

In a lattice regularization, the constant $\Lambda$ in Eq.~\eqref{eq:statenergyII} accounts for the linear divergence due to the self-energy.
While in dimensional regularization the linear divergence vanishes, the constant $\Lambda$ is still there, as it encodes a renormalon ambiguity of order $\LQCD$ that cancels against the renormalon ambiguity of the same order in the color-singlet static potential~\cite{Pineda:1998id, *Hoang:1998nz_short}.
The renormalon in the static potential is responsible for the poor convergence of its perturbative series.
The poor convergence may be cured by subtracting the renormalon contribution from the static potential in a suitably chosen renormalon subtraction scheme and reabsorbing it into a redefinition of $\Lambda$, making in this way the renormalon cancellation explicit~\cite{Pineda:2002se}.

Another way to get rid of the renormalon in the perturbative expansion of the static energy is by computing first the force,
\begin{equation}
F_{0}(r) \equiv \frac{\d}{\d r} E_{0}(r) ,
\end{equation}
which is free from the leading order renormalon and therefore well behaved as an expansion in $\als$.
One then recovers the static energy by integrating~\cite{Necco:2001gh}
\begin{equation}
\label{eq:V_from_F}
E_{0}(r) = \int\limits_{r^{*}}^{r} \d r' F_{0}(r') + \text{const} .
\end{equation}
The distance $r^{*}$, which should be smaller than $r$, is arbitrary, and contributes only with an additive constant.
This constant can be reabsorbed into an additive shift when comparing with lattice data.
Equation~\eqref{eq:V_from_F} effectively amounts to a rearrangement of the perturbative series that is renormalon free at order $\LQCD$~\cite{Pineda:2002se}.
The integral in Eq.~\eqref{eq:V_from_F} can be computed (numerically) while running the strong coupling with a soft scale that is chosen to be the inverse of the distance, $\nu = 1/r$.

A determination of the force on the lattice, which would allow for a direct comparison with its perturbative expression without managing the renormalon via subtraction or Eq.~\eqref{eq:V_from_F}, remains challenging.
On the one hand, full QCD data for the static energy at short distances are still too sparse to allow for very accurate determinations of the force from finite differences~\cite{Bazavov:2014soa_short}.
On the other hand, a direct computation of the force from Wilson loops with a chromoelectric field insertion seems to converge only slowly towards the continuum limit~\cite{Brambilla:2021wqs_short}.
For the time being, an accurate determination of $E_{0}(r)$ is much easier and precise, as it amounts to extracting the exponential fall-off of a static Wilson loop.

\subsection{Charm-quark effects in perturbation theory}
\label{subsec:charm_pert}

The static energy and potential are known at three loops only in the case of massless dynamical quarks.
Including a massive quark of mass $m$ to the static potential with $\Nf$ massless flavors yields the following correction to the energy
\begin{equation}
\label{eq:full_statenergy}
E_{0,m}^{(\Nf)}(r) = E_{0}^{(\Nf+1)}(r) + \delta V_{m}^{(\Nf+1)}(r) .
\end{equation}
The energy $E_{0}^{(\Nf+1)}(r)$ is the static quark-antiquark energy with $\Nf+1$ massless quarks.
The correction to it, $\delta V_{m}^{(\Nf+1)}$, due to one quark of finite mass $m$ is known at N$^{2}$LO.
The $\O(\als^{2})$ corrections~\cite{Billoire:1979ih} are, together with the $\O(\als^{3})$ corrections, well known~\cite{Melles:2000dq, *Eiras:2000rh, *Melles:2000ey, *Hoang:2000fm}; for a typo-free summary, see Ref.~\cite{Recksiegel:2001xq}.
Both, $E_{0}$, and $\delta V_{m}$, are to be evaluated with $\Nf+1$ massless flavors which, upon combination, will give the energy with $\Nf$ massless flavors and one massive one.
We reexpress Eq.~\eqref{eq:full_statenergy} in terms of the running coupling with $\Nf$ massless flavors.
On general grounds, the static energy with a massive quark satisfies the following limiting behavior:
\begin{enumerate}
\item In the limit $m \gg 1/r$, the heavy quark effectively decouples:
\begin{equation}
E_{0,m}^{(\Nf)}(r,\nu) \rightarrow E_{0}^{(\Nf)}(r,\nu) + \O((\als^{(\Nf)})^{4}) .
\end{equation}
\item In the limit $m \ll 1/r$, the static energy with a massive quark effectively reduces to the static energy with $\Nf+1$ massless quarks:
\begin{equation}
E_{0,m}^{(\Nf)}(r,\nu) \rightarrow E_{0}^{(\Nf+1)}(r,\nu) + \O((\als^{(\Nf)})^{4}) .
\end{equation}
\end{enumerate}

\section{Static energy and charm-quark effects on the lattice}

\subsection{Lattice setup and static quark-antiquark energy on the lattice}

We employ $2+1$-flavor ensembles~\cite{HotQCD:2014kol, Bazavov:2017dsy} from HotQCD and $2+1+1$-flavor ensembles~\cite{Bazavov:2012xda, Bazavov:2017lyh} from MILC, both using (rooted) highly improved staggered quark (HISQ) action~\cite{Follana:2006rc_short} for the sea quarks and, respectively, tree-level or 1-loop Symanzik improved gauge action.
Details of the ensembles can be found in the original references.
The static energy, $E_{0}(r)$, is extracted from the correlator of two Wilson lines in Coulomb gauge, and we extract the ground state by employing single and multi-exponential fits.
We always take the distance $r$ to be the distance inferred from the gluon propagator, which we refer to as tree-level improvement of the static-energy data.

Beyond tree-level improvement of the data, the $2+1$-flavor data are nonperturbatively improved~\cite{Bazavov:2019qoo_short}.
We use the single ensemble with $\beta = 7.825$, $a \approx 0.040$~fm, and $m_{l}/m_{s} \approx 1/20$, i.e., an almost physical quark-mass ratio.

A summary of the $2+1+1$-flavor ensembles can be found in Ref.~\cite{Bazavov:2017lyh}.
The data are tree-level improved.
In addition to the straight Wilson lines, we also determine the static energy from Wilson lines with one step of hypercubic (HYP) smearing~\cite{Hasenfratz:2001hp} which gives a better signal-to-noise ratio at large distances.
Although these two determinations of the static energy differ at nonzero lattice spacing, they share a common continuum limit.
In these proceedings, we mainly focus on the ensemble with $\beta = 7.0$, $a \approx 0.043$~fm, $N_{\sigma}^{3} \times N_{\tau} = 144^{3} \times 288$, and $m_{l}/m_{s} \approx 1/27$, which is the physical ratio.

\subsection{Charm-quark effects on the lattice}

\begin{figure}[t]
\centering
{\begin{overpic}[width=0.43\textwidth]{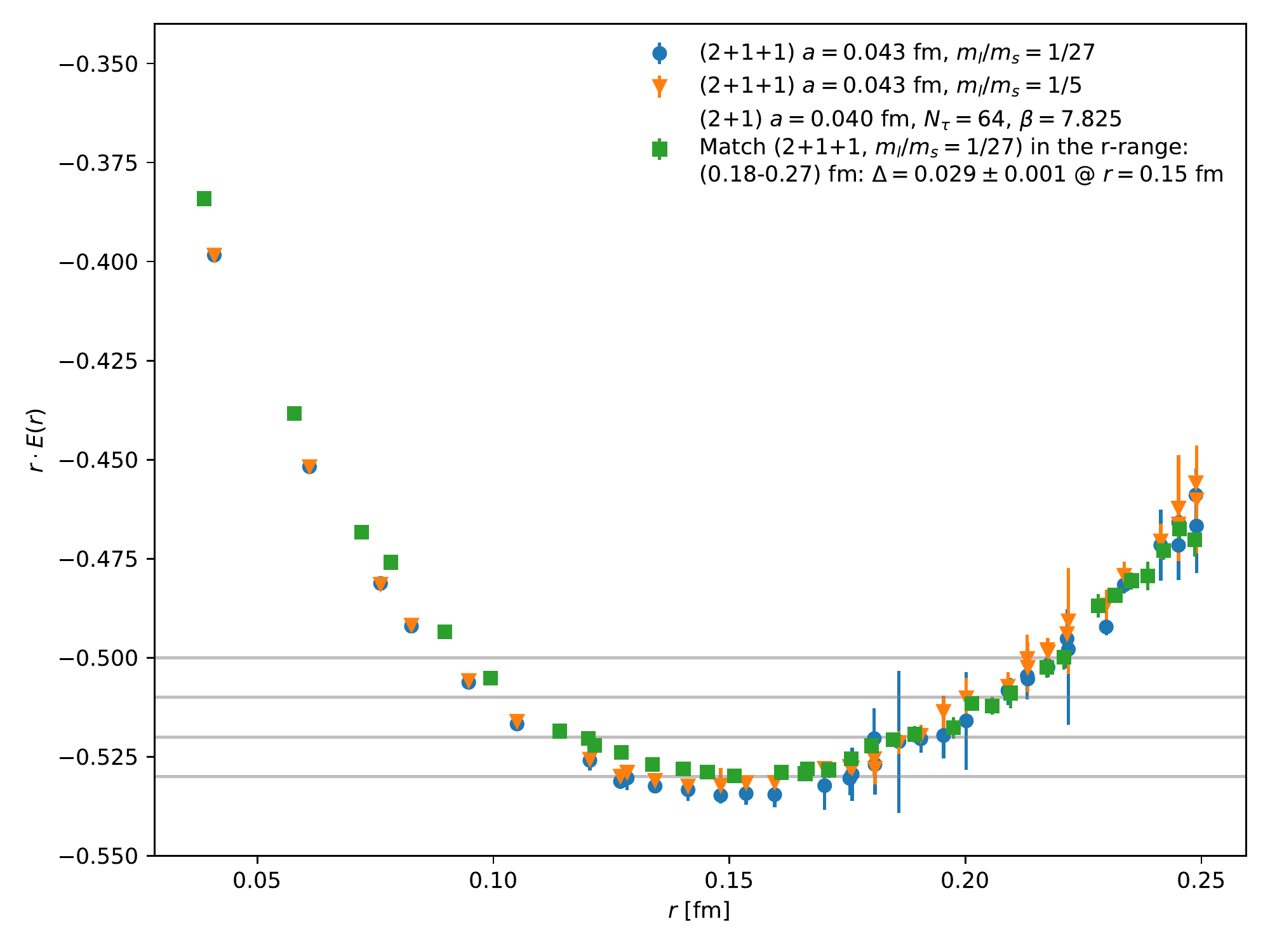}%
\put (15,40) {\hbox{\huge \color{gray} \transparent{0.4} PRELIMINARY}}
\end{overpic}}%
\hfill%
{\begin{overpic}[trim={0.4cm 0.5cm 32.6cm 1.4cm}, clip, width=0.50\textwidth]{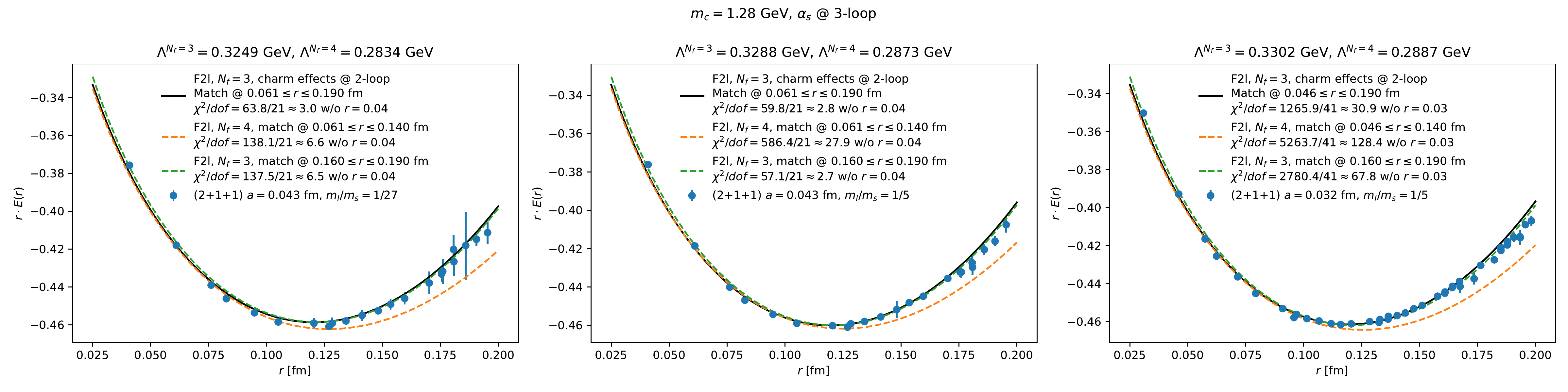}%
\put (20,25) {\hbox{\huge \color{gray} \transparent{0.4} PRELIMINARY}}
\end{overpic}}
\caption{\label{fig:charm}%
Left: $r  E_{0}(r)$ for two different $2+1+1$-flavor ensembles using different light-quark masses and one $2+1$-flavor ensemble of similar lattice spacing.
The latter has been matched to the $2+1+1$-flavor ensemble at large distances where they agree due to decoupling of the charm-quark while at distances $r \sim 1/\mc \sim 0.15$~fm the difference is due to the dynamic charm-quark.\newline
Right: Comparison of the $2+1+1$-flavor data with different perturbative curves describing the data in different regions (see text for explanation).
In black we show the two-loop static potential obtained from the static force ("F2l", see Eq.~\eqref{eq:V_from_F}) with three massless flavors ("$\Nf=3$") and a massive charm-quark at two loops (charm effects @ 2-loop), while in orange and green we show the pure two-loop massless four and three flavor curves respectively.}
\end{figure}

In the left panel of Fig.~\ref{fig:charm}, we show $rE_{0}(r)$, in order to remove the leading Coulomb behavior.
We add a constant to the $E_{0}(r)$ data such that the shifted data are rather flat in the range of interest.
This should facilitate the visualization of the small finite mass effects we are investigating.
Assuming decoupling of the charm-quark at large distances, $r \gg 1/\mc \sim 0.15$~fm, we match the $2+1$ and $2+1+1$-flavor data, since they must agree up to a constant.
We match the $2+1$-flavor data to the $2+1+1$-flavor data of the same (similar) $m_{l}/m_{s}$-ratio by minimizing the difference over the range $r \in [0.18,0.27]$~fm and vary the range to estimate the matching error.
In the legend of the plot, we display by how much, $\Delta$, the matching procedure shifts the data at $r = 0.15$~fm.
We additionally show another $2+1+1$-flavor data set with larger light sea quark mass, $m_{l}/m_{s} = 1/5$, whose data set has not been shifted relative to the physical one, such that their relative deviation is due to the difference in the light quark masses.
There is a clear difference between the data points of the $2+1$ and the $2+1+1$-flavor ensembles, which we attribute to the dynamical charm as the effect due to the different light-quark masses is smaller.

As discussed in Sec.~\ref{subsec:charm_pert}, the effective number of active flavors that enters the running of $\als$ in the static energy changes at different distances with $\mc$ fixed.
At large distances, $r \gg 1/\mc$, the charm-quark decouples, and in this region we thus have an effective three-flavor running of $\als$.
At short distances, $r \ll 1/\mc$, the charm-quark contributes as an active, massless flavor, and we thus have an effective four-flavor running of $\als$ in that region.
In order to compare with the weak-coupling prediction for charm-quark effects, Eq.~\eqref{eq:full_statenergy}, which is known at two-loop accuracy, we need a reference value for $\Lambda_{\MS}$.
We determine a sensible $\Lambda_{\MS}$ by fitting Eq.~\eqref{eq:full_statenergy} to the physical $2+1+1$-flavor ensemble.
We leave out data at $r/a=1$ from all the fits and vary the fit range up to $r \approx 0.18$~fm using $\mc = 1.28$~GeV and three-loop running of $\als$.
The value obtained after converting to three-flavors via perturbative decoupling~\cite{Chetyrkin:2000yt, *Schmidt:2012az, *Herren:2017osy}, $\Lambda_{\MS}^{(\Nf=3)} \approx 325$~MeV, turns out $\sim 3.5\%$ higher than the 2019 $2+1$-flavor determination from Ref.~\cite{Bazavov:2019qoo_short}, which is compatible within perturbative truncation uncertainties.

The comparison of the N$^{2}$LO expression with the lattice data is shown in the right panel of Fig.~\ref{fig:charm}.
The static energy, Eq.~\eqref{eq:statenergyII}, with four massless active flavors, $\Nf=4$, is matched to the data in the smaller $r$-range ($r < 1/\mc$, orange dashed curve), but begins to deviate from the lattice data before $r \approx 0.12$~fm.
Over the whole data range, $(0.061 - 0.190)$~fm, it has $\chi^{2}/\text{dof} \approx 6.6$.
The static energy, Eq.~\eqref{eq:statenergyII}, with three massless active flavors, $\Nf=3$, is matched to the data in the larger $r$-range ($r > 1/\mc$, green dashed curve).
It describes the data with a $\chi^{2}/\text{dof} \approx 6.5$, which is similar to the previous one for the $\Nf=4$ massless quark case.
We note, however, that with the value of $\Lambda_{\MS}^{(\Nf=3)}$ used here and with the choice $\nu=1/r$ for the soft scale, we are no longer able to describe the $2+1$-flavor data at short distances, and matching to the data at short distance would no longer describe the long distance data.
Finally, using three massless active flavors, $\Nf=3$, and a massive charm-quark, matched to the data over the whole $r$ range (black solid curve), we get a reasonable description of the data, and over the whole data range it has the best $\chi^{2}/\text{dof} \approx 3.0$.
The black curve also smoothly interpolates between the $\Nf=4$ massless quark case (at short distances) and the $\Nf=3$ massless quark case (at large distances).

\section{Lattice scales and the string tension}

The static energy based lattice scales $r_{i}/a$, $i=0,1,2$, have not been determined in $2+1+1$-flavor QCD, yet.
They are defined via ($\rho \equiv r/a$)~\cite{Bazavov:2017dsy, Sommer:1993ce, *Bernard:2000gd_short} 
\begin{equation}
\left. \rho^{2} F(\rho) \right|_{\rho = r_{i}/a} = \begin{cases}
\begin{array}{rl}
1.65 , & i=0 \\
1.0 , & i=1 \\
0.5 , & i=2
\end{array}
\end{cases} ,
\end{equation}
and have the continuum values~\cite{MILC:2010hzw, *Sommer:2014mea, HotQCD:2014kol, Bazavov:2017dsy}
\begin{equation}
r_{0} \approx 0.475~\text{fm} , \quad r_{1} \approx 0.3106~\text{fm} , \quad r_{2} \approx 0.145~\text{fm} \sim 1/\mc .
\end{equation}
Since the short distance scale $r_{2}$ is of the order of the inverse charm-quark mass, we may expect some effect on it due to the dynamic charm-quark.

A general fit Ansatz is given by the Cornell parametrization ($\rho \equiv r/a$)
\begin{equation}
\label{eq:Cornell}
aE_{0}(\rho) = -A/\rho + B + \sigma a^{2} \rho , \quad \Rightarrow \quad a^{2}F(\rho) = A/\rho^{2} + \sigma a^{2},
\end{equation}
where $\sigma$ corresponds to the string tension at distances that are large enough for confinement to set in, yet small enough to avoid string breaking.
The individual fits for the three scales and also for the string tension are realized as local fits in the region around the respective scales or at large distances only, respectively.
We estimate the covariance among the $E_{0}$ data at different $r$ by applying the jackknife method to the correlator fits, and we then propagate the errors through pseudorandom sampling of the resulting multivariate normal distribution.

For each $r_{i}$, we choose a range $\pm 15\%$ around the continuum value based on the $f_{p4s}$ scale, and demand at least three data points in the range.
This excludes determinations of $r_{2}/a$ on coarse ensembles.
Additionally, we thin out the data in order to be able to safely invert the correlation matrix by (i) equidistant or (ii) randomly choosing 10 points in the $\pm 15\%$ interval.
In the first method, we include the endpoints of the interval.

For the $\sigma$ determination, we vary $r_{\text{min}}$ around $0.6$~fm and use different values for the $A$-parameter.
We have not determined the Wilson line correlator at distances beyond $0.85$~fm.
The variation of $r_{\text{min}}$ uses the values $[0.58, 0.59, 0.60, 0.61]$~fm and the different $A$-values for the Cornell parametrization are $0$ (this assumes no Coulomb-like contribution), $A_{r_{0}}$, which is the $A$-parameter we get from the determination of the scale $r_{0}/a$ on the same sample (as described above), and, finally, $\pi/12$, which is the Lüscher term~\cite{Luscher:1980ac}.
We again thin out the data using both methods described above, but this time we keep 15 data points.

\begin{figure}[t]
\centering
{\begin{overpic}[page=11, trim={0.0cm 0.0cm 0.0cm 1.0cm}, clip, width=0.66\textwidth]{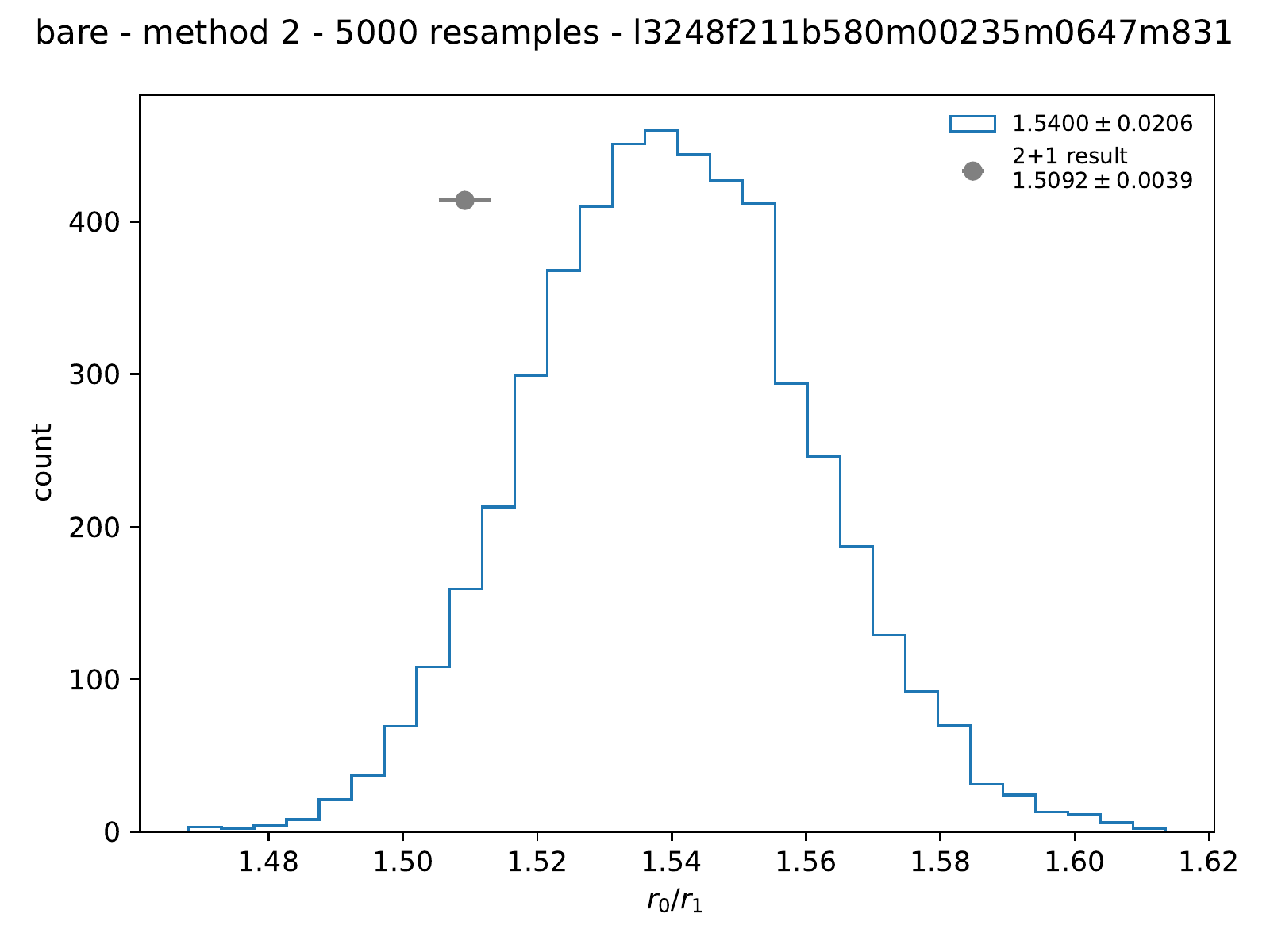}%
\put (10,15) {\hbox{\huge \color{gray} \transparent{0.4} PRELIMINARY}}
\end{overpic}}%
{\begin{overpic}[page=11, trim={0.3cm 0.5cm 32.5cm 12.7cm}, clip, width=0.32\textwidth]{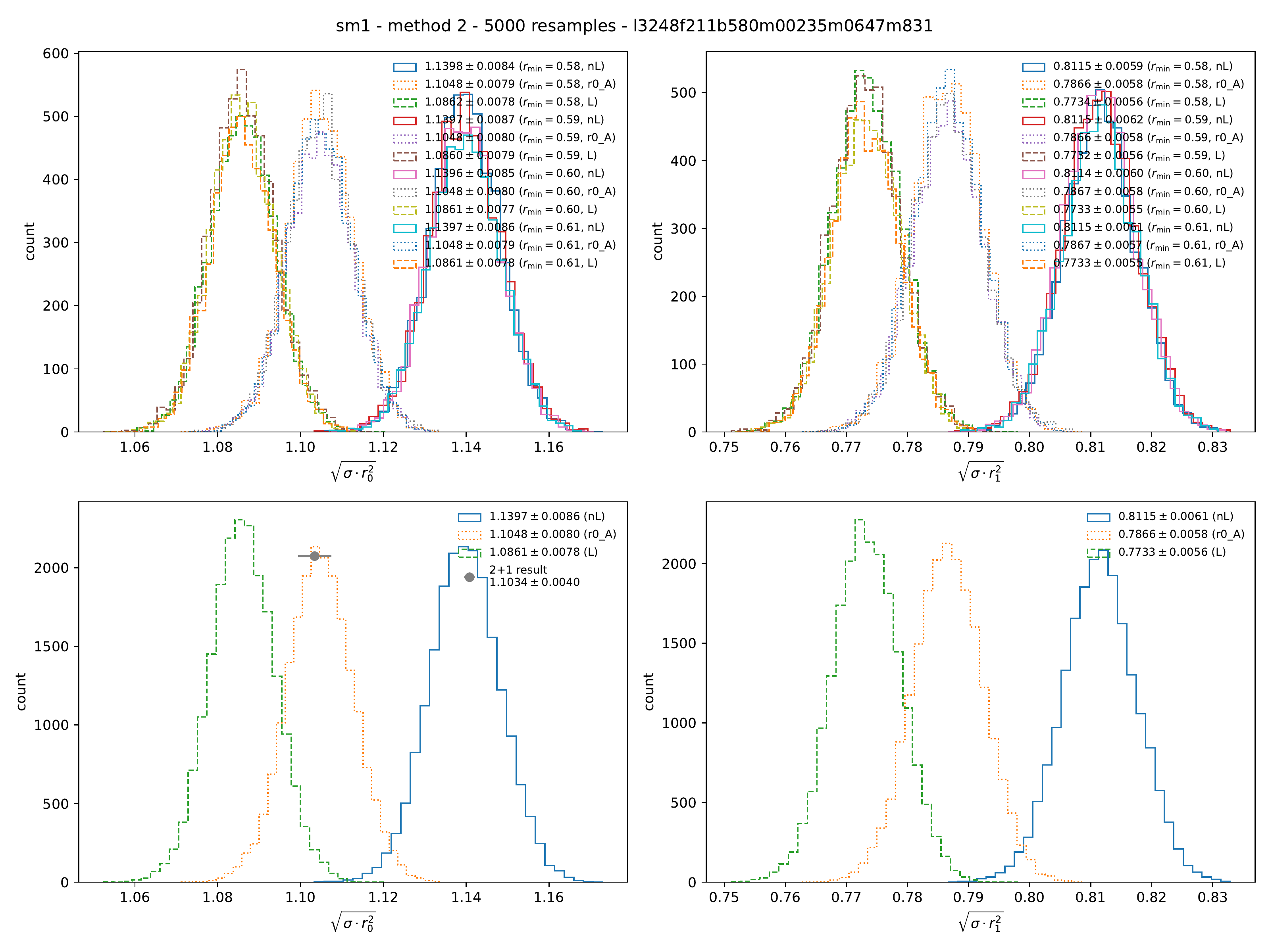}%
\put (-40,30) {\hbox{\huge \color{gray} \transparent{0.4} PRELIMINARY}}
\end{overpic}}
\caption{\label{fig:scales_sigma}%
Histograms of the sampling results for the quantities $r_{0}/r_{1}$ (left panel), $r_{1}/r_{2}$ (middle panel), and $\sqrt{\sigma r_{0}^{2}}$ (right panel) compared with the $2+1$-flavor results~\cite{HotQCD:2014kol, Bazavov:2017dsy, Cheng:2007jq} (gray points).
The three histograms of the string tension are for different choices of the $A$-parameter (see Eq.~\eqref{eq:Cornell}).}
\end{figure}

In Fig.~\ref{fig:scales_sigma}, we present histograms of 5000 samples thinning out the data by the second fitting method (see above) for the scale ratios $r_{0}/r_{1}$ and $r_{1}/r_{2}$ obtained from the bare result of the static energy, and the dimensionless string tension $\sqrt{\sigma  r_{0}^{2}}$ obtained from the static energy using one step of HYP-smearing.

Due to both scales being larger than the inverse charm-quark mass, the ratio $r_{0}/r_{1}$ should not be affected by the massive charm-quark.
Indeed, this is seen in the left panel when comparing to the previous $2+1$-flavor determination~\cite{HotQCD:2014kol}.
The good agreement is consistent with decoupling at large distances.
The ratio $r_{1}/r_{2}$, instead, should be affected by the massive charm-quark due to the scale $r_{2}$ being of the order of $1/\mc$.
Indeed, the result shown in the middle panel deviates from the continuum extrapolated $2+1$-flavor determination~\cite{Bazavov:2017dsy} by about 4--5\%, indicating an effect of the dynamical charm quark.

Finally, we determine the string tension for this ensemble using one step of HYP-smearing.
We see no significant variation as a function of $r_{\text{min}}$ and, therefore, we combine the four results that we get for each choice of $A$ into the one histogram in the right panel.
There is no significant difference between choosing the Lüscher term (green histogram) or $A_{r_{0}}$ (orange histogram) -- either are compatible with the determination from~\cite{Cheng:2007jq}, where data up to $1$~fm and the Lüscher term was used.
Neglecting the Coulomb-like term (blue histogram), however, gives a significantly larger result no longer compatible with $2+1$-flavors~\cite{Cheng:2007jq}.

We get consistent results for the scales, ratios, and string tension using both methods for choosing the points out of the fit ranges outlined above.

\section{Summary}

In summary, our lattice QCD data show effects of a nonzero charm-quark mass in the static energy.
Moreover, these effects are in agreement with perturbation theory at two-loop accuracy.
We stress that the static energy with massive charm contributions is known in perturbation theory one order in $\als$ less than the static energy with just massless quarks, which limits the accuracy of the analysis.
That said, the two-loop perturbative description with $\Nf=3$ massless and one massive flavor interpolates between the limiting $\Nf=4$ and $\Nf=3$ cases of the lattice data.
More specifically, we observe (i) the approach to the massless charm limit at short distances ($r \ll 1/\mc$), where the data can be described by the expression of the static energy with $\Nf=4$ massless quarks, and (ii) the decoupling of the massive charm quark in the long distance limit ($r \gg 1/\mc$), where the $2+1+1$ or $2+1$-flavor results are both consistent with the $\Nf=3$ result.

We get overall consistent estimates for the scales $r_{i}/a$ and of $\sqrt{\sigma  r_{i}^{2}}$ for the first time in $2+1+1$-flavor simulations.
Especially, the ratio $r_{0}/r_{1}$ and $\sqrt{\sigma  r_{0}^{2}}$ seem consistent with earlier $2+1$-flavor determinations in~\cite{HotQCD:2014kol, Cheng:2007jq} and thus are indicative of the decoupling of the charm quark.
The ratio $r_{1}/r_{2}$ on the other hand deviates from the $2+1$-flavor result~\cite{Bazavov:2017dsy} due to the effects from the massive charm quark on $r_{2} \sim 1/\mc$.
For the string tension, we can see some dependence on the choices of $A$ and $r_{\text{min}}$.

The results in these proceedings are still preliminary.
The uncertainties quoted so far are of purely statistical nature, because we focus on a single ensemble.
We are refining the ground state extraction in order to obtain a complete, solid error budget, and we are analyzing several of the $2+1+1$-flavor HISQ ensembles.


\acknowledgments
\noindent
We thank the MILC Collaboration for allowing use of their $2+1+1$-flavor HISQ ensembles.
The simulations were carried out on the computing facilities of the Computational Center for Particle and Astrophysics (C2PAP) in the project \emph{Calculation of finite $T$ QCD correlators} (pr83pu) and of the SuperMUC cluster at the Leibniz-Rechenzentrum (LRZ) in the project \emph{The role of the charm-quark for the QCD coupling constant} (pn56bo).
This research was funded by the Deutsche Forschungsgemeinschaft (DFG, German Research Foundation) cluster of excellence ``ORIGINS'' (\href{www.origins-cluster.de}{www.origins-cluster.de}) under Germany's Excellence Strategy EXC-2094-390783311.
Fermilab is managed by Fermi Research Alliance, LLC, under Contract No.\ DE-AC02-07CH11359 with the U.S.\ Department of Energy.
P.~Petreczky is supported by the U.S.\ Department of Energy under Contract No.\ DE-SC0012704.
R.~L.~Delgado is supported by the Ram{\'o}n Areces Foundation, the INFN post-doctoral fellowship AAOODGF-2019-0000329, and the Spanish grant MICINN: PID2019-108655GB-I00.
J.~H.~Weber’s research is also funded by the Deutsche Forschungsgemeinschaft (DFG, German Research Foundation) -- Projektnummer 417533893/GRK2575 ``Rethinking Quantum Field Theory''.
S.~Steinbeißer would like to thank Florian~M.\ Kaspar for discussion.


\bibliographystyle{apsrevM}
\bibliography{static_potential,\jobname}


\end{document}